\documentclass[manuscript,screen]{acmart}
\usepackage{graphicx, comment}
\usepackage{amsfonts}
\usepackage{bm}
\usepackage{color}
\usepackage{epstopdf}
\usepackage{epsfig}
\usepackage{verbatim}
\usepackage{lineno}
\usepackage{subcaption}
\usepackage[thicklines]{cancel}
\usepackage{url}
\usepackage{soul}
\usepackage{wrapfig}

\usepackage{tikz}
\usepackage{pgf-umlcd}
\usepackage[edges]{forest}
\usetikzlibrary{arrows.meta,shadows.blur}
\colorlet{myyellow}{yellow!50}
\usepackage{float}
\usepackage[ruled,vlined]{algorithm2e}
\usepackage{dblfloatfix}
\usepackage{graphicx,float}
\usetikzlibrary{shapes,arrows}
\usepackage{amsmath,bm,times}
\usepackage{pgfplots}
\usepackage{pgfplotstable}
\usepackage{subcaption}
\usepackage[export]{adjustbox}
\usepackage{bera}
\usepackage{listings,mdframed}
\usepackage{xcolor}
\usepackage{multirow}
\usepackage{tablefootnote}

\usepackage{pythonhighlight}
\definecolor{comment-text-color}{rgb}{0,0.8,0.6}
\lstdefinelanguage{json}{
    basicstyle=\normalfont\ttfamily\footnotesize,
    numbers=left,
    numberstyle=\scriptsize,
    stepnumber=1,
    numbersep=8pt,
    showstringspaces=false,
    breaklines=true,
    frame=lines,
    backgroundcolor=\color{background},
    literate=
     *{0}{{{\color{numb}0}}}{1}
      {1}{{{\color{numb}1}}}{1}
      {2}{{{\color{numb}2}}}{1}
      {3}{{{\color{numb}3}}}{1}
      {4}{{{\color{numb}4}}}{1}
      {5}{{{\color{numb}5}}}{1}
      {6}{{{\color{numb}6}}}{1}
      {7}{{{\color{numb}7}}}{1}
      {8}{{{\color{numb}8}}}{1}
      {9}{{{\color{numb}9}}}{1}
      {:}{{{\color{punct}{:}}}}{1}
      {,}{{{\color{punct}{,}}}}{1}
      {\{}{{{\color{delim}{\{}}}}{1}
      {\}}{{{\color{delim}{\}}}}}{1}
      {[}{{{\color{delim}{[}}}}{1}
      {]}{{{\color{delim}{]}}}}{1},
}

\usepackage{xcolor}
\usepackage{listings}
\usepackage{enumerate}
\usepackage{siunitx}
\usepackage{booktabs}
\usepackage{hyperref}
\hypersetup{
     colorlinks   = true,
     citecolor    = blue
}
\definecolor{red}{rgb}{1,0.,0}

\usepackage[compatibility=false]{caption}
\usepackage{tcolorbox}
\setcopyright{none}
\begin{document}

\title[Tensor Network Quantum Virtual Machine ]{Tensor Network Quantum Virtual Machine for Simulating Quantum Circuits at Exascale}

\thanks{This manuscript has been authored by UT-Battelle, LLC under Contract No. DE-AC05-00OR22725 with the U.S. Department of Energy. The United States Government retains and the publisher, by accepting the article for publication, acknowledges that the United States Government retains a non-exclusive, paid-up, irrevocable, world-wide license to publish or reproduce the published form of this manuscript, or allow others to do so, for United States Government purposes. The Department of Energy will provide public access to these results of federally sponsored research in accordance with the DOE Public Access Plan. (http://energy.gov/downloads/doe-public-access-plan).}

\begin{abstract}
The numerical simulation of quantum circuits is an indispensable tool for development, verification and validation of hybrid quantum-classical algorithms on near-term quantum co-processors. The emergence of exascale high-performance computing (HPC) platforms presents new opportunities for pushing the boundaries of quantum circuit simulation. We present a modernized version of the Tensor Network Quantum Virtual Machine (TNQVM) which serves as a quantum circuit simulation backend in the eXtreme-scale ACCelerator (XACC) framework. The new version is based on the general purpose, scalable tensor network processing library, ExaTN, and provides multiple configurable quantum circuit simulators enabling either exact quantum circuit simulation via the full tensor network contraction, or approximate quantum state representations via suitable tensor factorizations. Upon necessity, stochastic noise modeling from real quantum processors is incorporated into the simulations by modeling quantum channels with Kraus tensors. By combining the portable XACC quantum programming frontend and the scalable ExaTN numerical backend we introduce an end-to-end virtual quantum development environment which can scale from laptops to future exascale platforms. We report initial benchmarks of our framework which include a demonstration of the distributed execution, incorporation of quantum decoherence models, and simulation of the random quantum circuits used for the certification of quantum supremacy on the Google Sycamore superconducting architecture.
\end{abstract}

\author{Thien Nguyen}
\affiliation{
\institution{Quantum Computing Institute,  Oak Ridge National Laboratory}
\city{Oak Ridge, TN}
\country{USA}}
\affiliation{
\institution{Computer Science and Mathematics Division, Oak Ridge National Laboratory}
\city{Oak Ridge, TN}
\country{USA}}

\author{Dmitry Lyakh}
\affiliation{
\institution{Quantum Computing Institute,  Oak Ridge National Laboratory}
\city{Oak Ridge, TN}
\country{USA}}
\affiliation{
\institution{National Center for Computational Sciences, Oak Ridge National Laboratory}
\city{Oak Ridge, TN}
\country{USA}}

\author{Eugene Dumitrescu}
\affiliation{
\institution{Quantum Computing Institute,  Oak Ridge National Laboratory}
\city{Oak Ridge, TN}
\country{USA}}
\affiliation{
\institution{Computational Sciences and Engineering Division, Oak Ridge National Laboratory}
\city{Oak Ridge, TN}
\country{USA}}

\author{David Clark}
\affiliation{
\institution{NVIDIA Corp.}
\city{Santa Clara, CA}
\country{USA}}

\author{Jeff Larkin}
\affiliation{
\institution{NVIDIA Corp.}
\city{Santa Clara, CA}
\country{USA}}

\author{Alexander McCaskey}
\email{mccaskeyaj@ornl.gov}
\affiliation{
\institution{Quantum Computing Institute,  Oak Ridge National Laboratory}
\city{Oak Ridge, TN}
\country{USA}}
\affiliation{
\institution{Computer Science and Mathematics Division, Oak Ridge National Laboratory}
\city{Oak Ridge, TN}
\country{USA}}

\maketitle
%%%%%%%%%%%%%%%%%%%%%%%%%
\section{Introduction}
\par
Quantum circuit simulation on classical computers is an important tool for development, verification, validation, and analysis of quantum algorithms in the noisy intermediate-scale quantum (NISQ) regime~\cite{Preskill_2018}. There exist a wide variety of simulation techniques that have been developed for this purpose, ranging from the state vector~\cite{H_ner_2017, DERAEDT201947, Guerreschi_2020} or density matrix~\cite{li2020density} simulators to Clifford-based~\cite{PhysRevA.70.052328, PhysRevLett.116.250501, PhysRevA.95.062337, Bravyi2019simulationofquantum, gidney2021stim} and tensor-based simulators~\cite{MarkovShi, mccaskey2018validating, Gray_Cotengra, Alibaba2020, Stoudenmire2020}. In particular, the tensor network based techniques have proven their power in constructing effective simulators of rather large quantum circuits with memory requirements that scale in accordance with the quantum state entanglement properties \cite{mccaskey2018validating, Stoudenmire2020}. More generally, tensor processing has been recognized as a computing technique applicable to many scientific and engineering domains~\cite{PhysRevA.74.022320, PhysRevX.8.031012, roberts2019tensornetwork, glasser2019probabilistic} that has resulted in highly-optimized software leveraging the state-of-the-art classical hardware capabilities to simulate complex physical phenomena \cite{Paolo2021}.

\par
The tensor network quantum virtual machine (TNQVM) was first introduced in \cite{mccaskey2018validating} as a tensor-based quantum circuit simulation back-end for the XACC framework~\cite{mccaskey2020xacc}. The original implementation leveraged the matrix product state (MPS) representation of the quantum circuit wave-function based on the data structures provided by the ITensor library~\cite{fishman2020itensor} --- a popular library that (at the time of implementation) only supported single-node CPU execution. In this work, we present an enhanced TNQVM implementation with a direct focus on HPC deployment via the utilization of the state-of-the-art Exascale Tensor Networks (ExaTN) library~\cite{exatnGithub} as the computational backend. This re-architected TNQVM code runs on both CPU and GPU hardware (including NVIDIA and AMD GPU), and supports multi-node, multi-GPU execution contexts. One of the primary drivers of this work is the need for a flexible high-performance simulator that can (1) extract experimentally verifiable results from large quantum circuits, and (2) take full advantage of computing resources by devising custom strategies for each simulation task and balance the workload (memory and compute) across all available resources.

Tensor network theory is a natural fit for large-scale quantum circuit simulations. Quantum computers encode computation and information in an exponentially large tensor space which is not directly accessible experimentally. One can only collect discrete observable statistics on a given quantum state (qubit measurement bit-strings, expectations values, etc.). The tensor network theory provides the most natural way of dealing with the exact and approximate tensor representations in such exponentially large spaces, thereby enabling an efficient expression of the quantum state observable quantities. Combined with a low-rank compression via low-order tensor factorizations, this approach also becomes highly amenable to memory-bound flop-oriented compute platforms, which most of the current HPC systems are. Thus, the main goal of the TNQVM code is to provide an implementation of a set of tensor network algorithms which are well suited for simulating quantum circuits in different use case scenarios. All necessary construction, manipulation, and processing of the derived tensor networks is automated via the ExaTN backend. Importantly, the ExaTN backend also provides the foundation for the simulation workload optimization. Not only can we describe quantum circuits in various tensor forms, such as matrix product state (MPS)~\cite{ORUS2014117}, tensor tree network (TTN)~\cite{PhysRevA.74.022320}, etc., but we can also delegate the runtime execution optimization to ExaTN where it can decompose and schedule the tensor processing tasks across all available resources, including multi-core CPUs, GPUs, and potentially more specialized accelerators.

In this manuscript, we present implementation details and demonstration results for the following TNQVM capabilities:
\begin{itemize}
    \item A generic tensor network contraction based simulator that expresses and evaluates the entire quantum circuit as an interconnected tensor network.
    \item A distributed-memory MPS-factorized state-vector simulator.
    \item A density matrix (noisy) simulator based on the hierarchical tensor network or locally-purified matrix product state representations.
    \item An automatic divide-and-conquer tensor network reconstruction simulator which synthesizes the optimal tensor network representations dynamically.
\end{itemize}
\par
We note that the generic tensor network contraction based simulator supports both noiseless and noisy simulations --- we use tensor networks to represent either the state vector or density matrix evolution, respectively. The noise-modelling operations expressed in terms of the channel (Kraus) operators can be incorporated into the latter to mimic the hardware noise models. Similarly, approximate tensor representations of pure and mixed quantum states, based on different tensor factorizations, are provided as alternative approaches geared towards larger-scale simulations. In essence, TNQVM provides a multi-modal simulation platform whereby one can quickly prototype and evaluate accuracy, runtime, parallelism, memory consumption, etc., of varying tensor-based approaches for quantum circuit simulation, as well as execute the actual production runs on workstations, HPC platforms and clouds.

\par
Compared to many other available quantum circuit simulation platforms, the XACC-TNQVM-ExaTN software stack offers unique features in terms of scalability, flexibility, extensibility, performance, and availability. There are not many quantum circuit simulators that have been rigorously tested in a state-of-the-art HPC environment. For example, the Flexible Quantum Circuit (qFlex) Simulator~\cite{Villalonga_2019, Villalonga_2020} and the QCMPS simulator~\cite{Dang2019} have demonstrated scalability and accuracy on large supercomputers for the contraction-based and MPS simulation approaches, respectively. However, both of these simulators have been developed for rather specific and narrow use cases, not targeting generic quantum circuit simulation workflows in a complete end-to-end quantum programming stack. Moreover, most tensor-based simulators, including qFlex and QCMPS, tend to associate their internal representation to a particular form of tensor networks as opposed to the multi-modal flexibility of TNQVM. Very recently, classical simulations of the random quantum circuits used in the Google quantum supremacy experiments~\cite{Arute_2019} have been revisited with new simulators after discovering a powerful tensor contraction path optimization algorithm~\cite{Gray_Cotengra}. The QUIMB~\cite{QUIMB} and ACQDP~\cite{Alibaba2020} simulators have positioned themselves as potentially the fastest simulators for the quantum supremacy circuits on distributed HPC platforms and clouds. Although they can be used for simulating other quantum computing circuits as well, their main focus has so far been on the direct tensor network contraction technique and noiseless amplitudes.

Last but not least, we note that the modular full-stack integration between XACC, TNQVM, and ExaTN allows us to support multiple quantum programming languages and run on different classical compute platforms seamlessly. This full-stack integration proves beneficial, especially for the TNQVM noisy simulators, which can query device noise models directly from the cloud-based hardware providers, e.g., IBM, using the XACC remote connection capabilities. Finally, we also want to stress our commitment to open-source development principles. All of our development activities and implementations are in the public domain under permissive licenses. 

The subsequent sections are organized as follows. Section~\ref{sec:background} provides some background information about the XACC programming framework, which TNQVM extends, and the ExaTN library which provides the scalability and numerical backbone for TNQVM. Section~\ref{sec:tnqvm_impl} details the implementation of various simulators in TNQVM in terms of the tensor language used by ExaTN. Section~\ref{sec:demo} provides examples and demonstration results of TNQVM for various tasks ranging from a large-scale quantum circuit simulation to noisy quantum circuit modeling. Conclusions and outlooks are given in Section~\ref{sec:conclusion}.
\section{Background}
\label{sec:background}
\subsection{ExaTN library}
\label{sec:exatn_intro}
The ExaTN library (Exascale Tensor Networks) provides generic capabilities for construction, manipulation and processing of arbitrary tensor networks on single workstations, commodity clusters and leadership supercomputers~\cite{exatnGithub}. On heterogeneous platforms, it can leverage GPU acceleration provided by NVIDIA and AMD GPU cards. Our TNQVM simulator uses the ExaTN library as a parallel tensor processing backend. The native ExaTN C++ application programming interface (API) consists of declarative and executive API (Pybind11~\cite{pybind11} bindings are also available for Python users). The declarative API provides functions for constructing arbitrary tensors and tensor networks and performing different formal manipulations on them. The executive API provides functions for allocating tensor storage and parallel processing of tensor networks, for example to perform tensor network contraction. The latter operation is automatically decomposed by the ExaTN parallel runtime into smaller tasks which are distributed across all MPI processes. 

ExaTN also provides API for higher-level algorithms, specifically for tensor network reconstruction and tensor network optimization. Tensor network reconstruction allows approximation of a given tensor network by another tensor network, normally with a simpler structure. Tensor network optimization allows finding extrema of a given symmetric tensor network functional (expectation value of some tensor operator). The provided capabilities are sufficient for reformulating generic linear algebra procedures to be restricted to low-rank tensor network manifolds. Such a low-rank compression of linear algebra procedures allows their efficient computation for rather large problems with a tiny fraction of their exact Flop and memory cost.

\subsection{XACC quantum programming framework}
\label{sec:xacc_intro}
XACC is a system-level quantum programming framework that enables language-agnostic programming targeting multiple physical and virtual quantum backends via a novel quantum intermediate representation (IR). Ultimately, XACC puts forward a service-oriented architecture and defines a number of interfaces or extension points that span the typical quantum-classical programming, compilation, and execution workflow. This platform provides an extensible backend interface for quantum program execution in a retargetable fashion, and this is the interface we target for this work. TNQVM directly extends this layer and enables execution of the XACC IR via tensor-network simulation in a multi-model fashion. 

%TNQVM is a simulator QPU implementation for XACC, which is a system-level quantum programming framework. XACC is a C++-based infrastructure for language- and hardware- agnostic quantum programming, compilation, and execution. Having a service-oriented architecture, the framework allows for extensibility via standardized service interfaces. 

Here, we briefly summarize pertinent XACC interfaces that are relevant to TNQVM and direct interested readers to~\cite{mccaskey2020xacc} for a comprehensive introduction. At the high-level, we can classify the framework components into three categories, namely frontend, middle-end, and backend. The frontend exposes a \texttt{Compiler} interface which is responsible for converting the input kernel source strings to the XACC \texttt{IR}. \texttt{IR} is a pertinent data-structure of the framework, capturing both the \texttt{Instruction} and \texttt{CompositeInstruction} service interfaces specialized for concrete quantum gates and collections of those gates, respectively. Using the IR representation of the quantum kernel as its core data structure, the middle-end layer also exposes an \texttt{IRTransformation} interface enabling general transformation of quantum circuits (\texttt{CompositeInstruction}) for tasks such as circuit optimization and qubit placement. Lastly, XACC provides an \texttt{Accelerator} interface enabling integration with physical and virtual (simulator) quantum computing backends. TNQVM implements this \texttt{Accelerator} interface, thus providing a universal quantum backend for the framework. In other words, one can use TNQVM interchangeably with other physical QPUs or simulators available in XACC. 

Internally, XACC \texttt{Accelerator} implementations usually leverage a visitor pattern (the XACC \\\texttt{InstructionVisitor}) \cite{gof} to walk the \texttt{IR} tree representation of compiled quantum circuits. Each \texttt{Accelerator} may opt to perform different actions while walking the \texttt{IR} tree. For instance, for physical hardware backends, the \texttt{Accelerator} adapter needs to convert XACC \texttt{IR} to the native gate set that the platform supports. As we will describe in detail later in the text, TNQVM makes use of this \texttt{InstructionVisitor} interface to construct different tensor network representations of the input circuit depending on the selected mode of simulation, e.g., exact or approximate, ideal or noisy simulation, etc.

\section{ExaTN-Enabled TNQVM Implementation}
\label{sec:tnqvm_impl}
The ultimate goal of our updated TNQVM implementation targeting leadership-class HPC systems is to map the input XACC IR to unique tensor data structure instances provided by ExaTN via the XACC \texttt{InstructionVisitor}. TNQVM will walk the IR tree via custom visitors and visit 
%In the full-stack XACC framework, TNQVM implements the \texttt{Accelerator} sub-type, taking as an input the compiled quantum circuit representation (\texttt{IR}). When walking this syntax tree using the ubiquitous visitor pattern, i.e. traversing 
the IR nodes (quantum gates) and construct, evaluate, and post-process corresponding ExaTN tensor and tensor network objects. Broadly speaking, TNQVM simulation methods can be categorized as either exact or approximate. The first category comprises backend simulators (visitors) that faithfully translate quantum circuits to fully-connected tensor networks and then contract them to evaluate the value of interest. On the other hand, approximate simulation methods rely on factorized forms of the state vector or density matrix where some form of a tensor network compression is used, for example, via the matrix product-state (MPS) tensor network. The factorized representation is maintained throughout the circuit simulation via a suitable decomposition procedure. Thus, we can balance the accuracy and complexity of these approximate representations throughout the simulation process. We note that TNQVM can incorporate stochastic noise into both forms of simulation.

\subsection{Direct Tensor Network Contraction}
\label{sec:direct_contraction}
\begin{figure}[h!] 
\centering
\begin{subfigure}{.25\textwidth}
  \centering
  \includegraphics[width=0.8\textwidth]{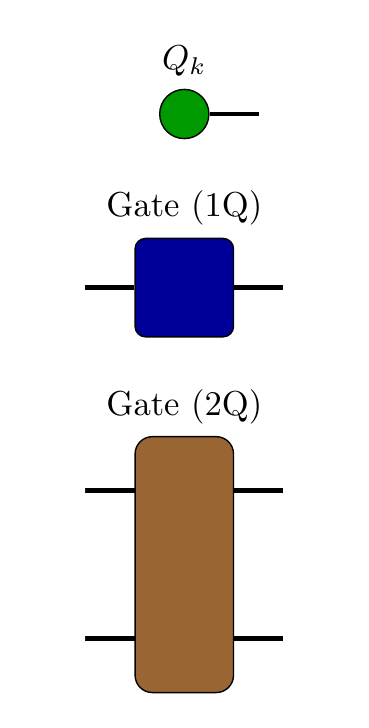} 
  \caption{Tensor notations for qubits, single and two-qubit quantum gates.}
  \label{fig:tensor_notations}
\end{subfigure}%
\begin{subfigure}{.75\textwidth}
  \centering
  \includegraphics[width=0.8\textwidth]{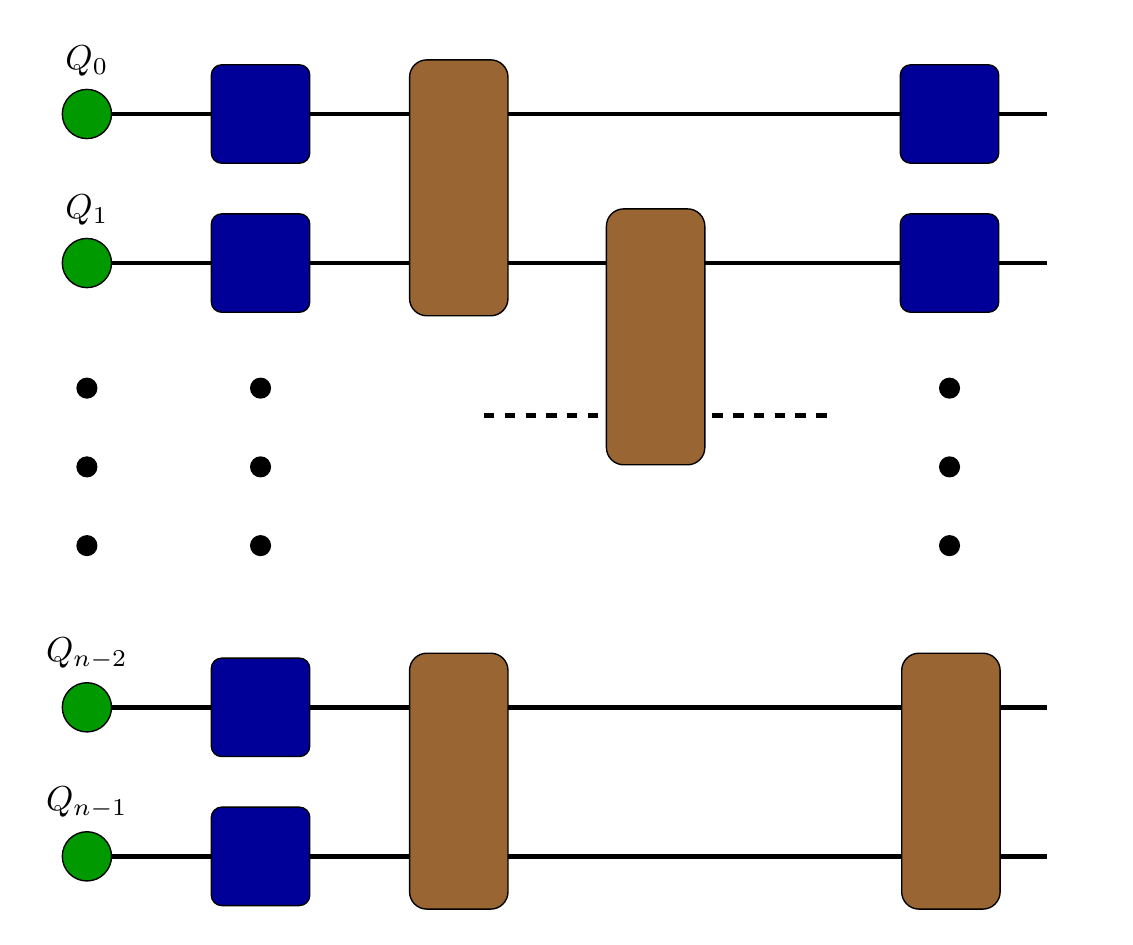} 
  \caption{Tensor network representation of a quantum circuit}
  \label{fig:circuit_tensor_network}
\end{subfigure}
\caption{ExaTN tensor network representation of a quantum circuit.}
\label{fig:exatn_circuit_tn}
\end{figure}

In this mode of execution, we construct a tensor network that embodies the entire quantum circuit before evaluating it numerically. More specifically, qubits, single-qubit gates, and two-qubit gates are represented by rank-1, rank-2, and rank-4 tensors, respectively, as depicted in Fig.~\ref{fig:tensor_notations}. 
The quantum circuit dictates the connectivity of tensors within the tensor network (see Fig.~\ref{fig:circuit_tensor_network}). Once completed, the tensor network is submitted to the ExaTN numerical server for parallel evaluation.

During the evaluation phase, ExaTN first analyzes the structure of the tensor network in order to determine the pseudo-optimal tensor contraction sequence, that is, it performs minimization of the Flop count necessary for the evaluation of the tensor network. The Flop count is minimized by using an algorithm based on recursive graph partitioning (via METIS library~\cite{Metis}) and some heuristics, following an approach similar to the one presented in Ref.~\cite{Gray_Cotengra}. Although the simpler optimizer implemented in ExaTN does not yet provide the same quality, it prioritizes the speed of the tensor contraction sequence search, to ensure that the search process does not take more time than the actual evaluation of the tensor network. Once a pseudo-optimal tensor contraction path has been found, the ExaTN numerical server executes the determined tensor contraction sequence across multiple nodes with an optional GPU acceleration capability. Each compute node executes its own subset of tensor sub-networks generated by slicing some of the tensor network indices, which also reduces the memory footprint of intermediate tensors. It is worth noting that ExaTN supports GPU processing of large tensors that do not fit into GPU memory. In this case, all cross-device data transfers are orchestrated by the library automatically and transparently to the user.

The direct tensor contraction works best for computing individual amplitudes or their batches. Since the number of open edges in the tensor network is equal to the number of qubits, we cannot obtain the full wave-function for a large number of qubits. Instead, for large-scale circuits, we have implemented a variety of utility functions to extract observable values, as described in Table~\ref{table:exatn_utils}.
\begin{table}
\caption{Tensor Network utility functions for evaluating observables for large-scale quantum circuits.}
\begin{center}
\begin{tabular}{ |p{0.16\textwidth}|p{0.7\textwidth}| } 
 \hline
 Mode & Description \\
 \hline
 Single-state amplitudes & Closing the quantum circuit tensor network with a $\langle \Psi_0 |$, where $\Psi_0$ represents a chosen bit-string. \\
 \hline
 Expectation value by conjugate  & Adding a tensor network which represents the observable and then closing with the conjugate quantum circuit. \\
 \hline
 Expectation value by state vector slicing  & A subset of open tensor legs is projected to a bitstring to keep the number of open legs within the memory constraints. Accumulating the expectation values computed on the partial state vector slices for all possible projected bitstrings to compute the overall expectation value. \\
 \hline
 Direct unbiased bit-string sampling & Connecting the quantum circuit tensor network with its conjugate while leaving a subset of qubit legs open to compute the reduced density matrices for bit-string sampling and measurement projection.\\
 \hline
\end{tabular}
\end{center}
\label{table:exatn_utils}
\end{table}

\subsubsection{Single-state amplitudes}
\begin{figure}[h!]
\centering
\includegraphics[width=0.8\textwidth]{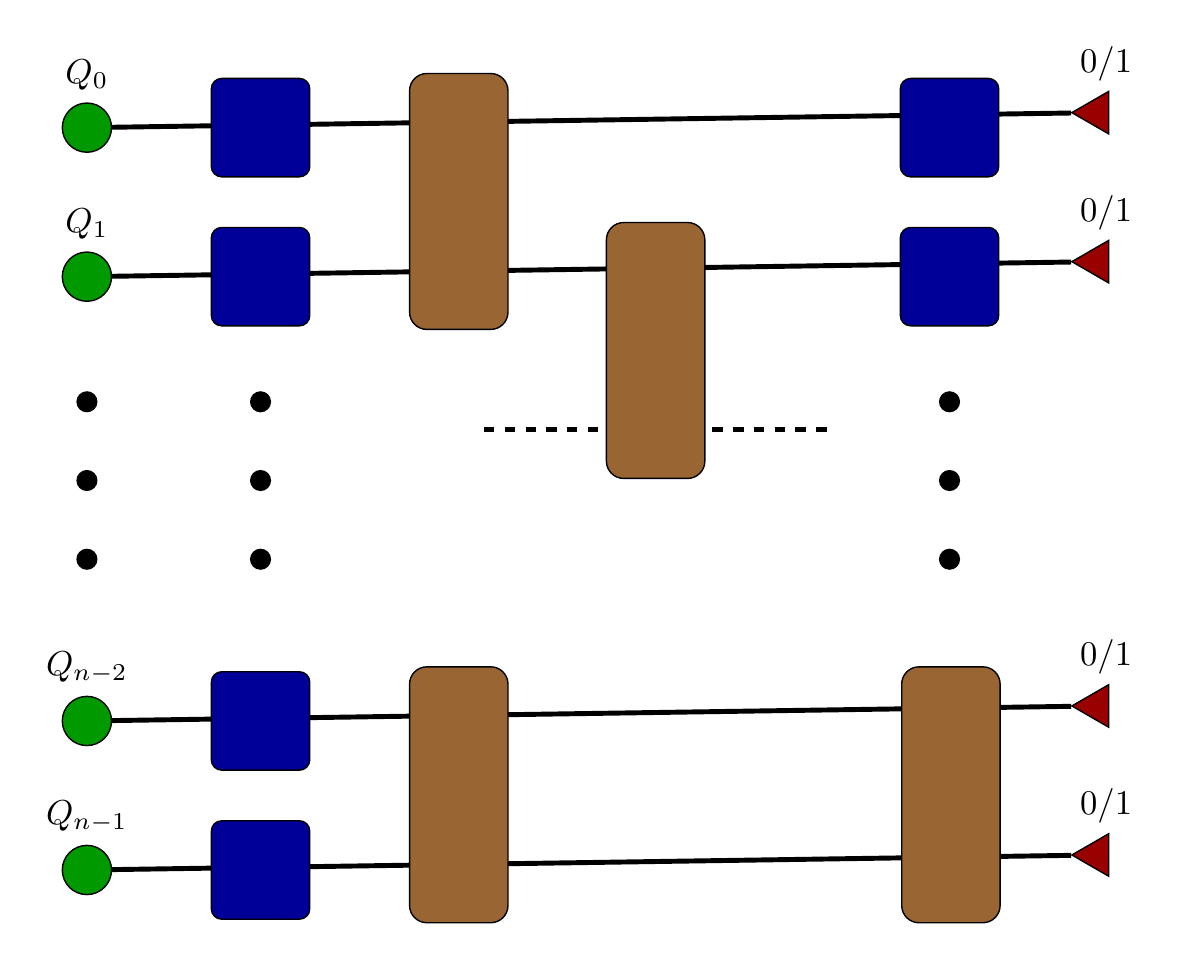} 
\caption{Single amplitude calculation by tensor network contraction: open qubit legs are closed with tensors representing the projected 0 or 1 values.}
\label{fig:ampl_calc}
\end{figure}

Once the full circuit tensor network has been constructed, we can append appropriate conjugate qubit tensors to each open qubit leg to compute a desired quantum state amplitude (as shown in Fig.~\ref{fig:ampl_calc}, red triangles project open qubit legs to a specific bit-string).

Effectively, here we construct the following tensor network to evaluate
\begin{equation}
\langle{\Psi_f}| U_{circuit} |\Psi_0\rangle,
\label{eq:ampl_calc}
\end{equation}
where $|\Psi_0\rangle$ and $U_{circuit}$ are the initial state and the equivalent unitary matrix of the quantum circuit, respectively. $|\Psi_f\rangle$ is the state whose amplitude we want to calculate. The result of~(\ref{eq:ampl_calc}) is just a scalar. This procedure can be repeated for different amplitudes. It is worth noting that despite its low rank, the numerical evaluation of a single-state amplitude for large-scale quantum circuits involving many qubits and gates is numerically challenging. Small gate tensors are contracted internally to form larger tensors and any intermediate tensors that require more memory than available will be split into smaller slices and distributed across multiple MPI processes. In principle, this allows us to simulate the output amplitudes of an arbitrarily large quantum circuit in terms of the number of qubits involved. This amplitude calculation plays a key role in validating near-term quantum hardware via procedures such as the random quantum circuit sampling protocol~\cite{Arute_2019}.

\subsubsection{Operator expectation values}
A ubiquitous use case in quantum circuit simulation is the calculation of the expectation values of hermitian operators, i.e.,
\begin{equation}
\langle{\Psi_f}| H |\Psi_f\rangle,
\label{eq:exp_val}
\end{equation}
where $|\Psi_f\rangle = U_{circuit} |\Psi_0\rangle$ is the final state of the qubit register and $H$ is a general hermitian operator representing an observable of interest. For instance, $H$ could be a hermitian sum of products of Pauli operators, $\{\sigma_I, \sigma_X, \sigma_Y, \sigma_Z\}$, on different qubits. We have developed two different methods to compute (\ref{eq:exp_val}) for circuits that have many qubits: (a) via the use of the conjugate tensor network, and (b) via the wave-function slicing approaches, as described in Table~\ref{table:exatn_utils}.

\begin{figure}[h!] 
\centering
\includegraphics[width=\textwidth]{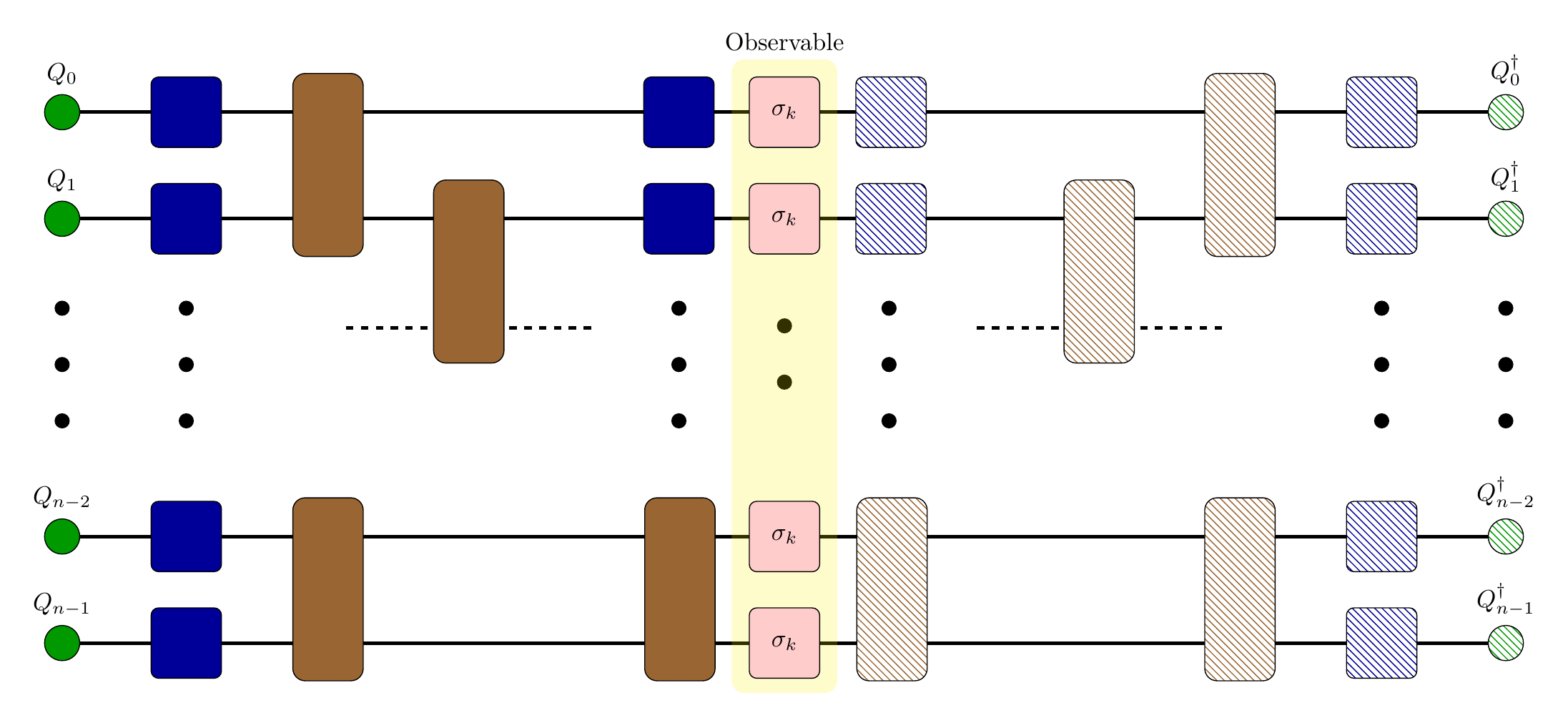} 
\caption{Expectation value calculation by a double depth circuit. The observable Pauli tensors $\{\sigma_k\} = \{ I, X, Y, Z \}$. Hatched tensors after observable Pauli operators are the conjugates of the ones on the left-hand side.}
\label{fig:exp_val_double_depth}
\end{figure}

In the first approach, after constructing the tensor network which represents $U_{circuit} |\Psi_0\rangle$, we append measure operator tensors and then the conjugate of the $U_{circuit} |\Psi_0\rangle$ network. The resulting tensor network evaluates to a scalar and consists of approximately twice the number of component tensors, as shown in Fig.~\ref{fig:exp_val_double_depth}.
The evaluation of this tensor network will give us the expectation value. It is worth mentioning that ExaTN can intelligently collapse a tensor and its conjugate in case they are contracted with each other, which can therefore simplify the tensor network if the measurement operator is sparse, i.e., involving only a small fraction of qubits in the circuit.

In the wave-function slicing evaluation method, we slice the output wave-function based on the memory constraint, compute the expectation value for each slice, and recombine them to form the final result at the end. Specifically, the workflow is as follows
\begin{enumerate}
    \item Based on the memory constraint, determine the max number of open qubit legs ($rank\_max$) in the output tensor.
    \item Determine the number of wave-function slices ($N_{slices}$), which is $2^{N_{projected}}$, $N_{projected} = N_{qubits} - rank\_max$.
    \item Distribute the wave-function slice compute tasks ($N_{slices}$) across all MPI processes.
    \item Compute the expectation value of the measurement operator for each slice.
    \item Sum (reduce) the partial expectation values to compute the final expectation value.
\end{enumerate}

As compared to the circuit conjugation method (double depth circuit), this approach has a couple of advantages. First, it does not require evaluation of a double-size tensor network. Second, in some scenarios, this slicing strategy can result in a more optimal workload. Third, we can distribute the partial expectation value calculation tasks in a massively parallel manner, which is amenable to large HPC platforms.

\subsection{Matrix Product State Simulation}
\label{sec:mps}
\par
TNQVM also provides an approximate simulator based on the MPS factorization of the circuit wave-function, where a user can specify the numerical limit for the singular value truncation as well as the maximum entanglement bond dimension. Built upon the parallel capabilities of ExaTN, we have implemented a distributed MPS tensor processing scheme in which the MPS tensors are distributed evenly across available compute nodes.

%The matrix product state (MPS) formalism provides a method to adapt the simulation numerical accuracy according to computational resource requirements. Specifically, the rank-n state vector tensor is decomposed into a tensor product train. We utilize \texttt{exatn::numerics::NetworkBuilder} utility to construct the initial MPS tensor network.

A quantum circuit simulation via the MPS simulator backend (named \texttt{exatn-mps}) is performed via the sequential contraction and decomposition steps. Single-qubit gate tensors can be absorbed into the qubit MPS tensors directly. The application of the two-qubit entangling gates on two neighboring MPS tensors is computed by:
\begin{itemize}
    \item Merge and contract the two MPS tensors with the rank-4 gate tensor.
    \item Decompose the resulting tensor back into two MPS tensors via the singular value decomposition API (\texttt{decomposeTensorSVDLR}) of ExaTN.
    \item Truncate the dimension of the connecting leg between the two post-SVD MPS tensors, the so-called bond dimension, according to chosen numerical accuracy or memory constraint settings.
    \item Update the MPS ansatz with the reduced-dimension MPS tensors.
\end{itemize}

By following this procedure, we can compute the MPS tensor network approximating the full state vector at the end of the quantum circuit. Expectation values or bit-string amplitudes can be computed in the same manner as previously described for the full tensor network contraction strategy. We also want to note that the transformation of quantum circuits into this nearest-neighbor form by injecting \texttt{SWAP} gates is performed automatically by the XACC \texttt{IRTransformation} service when the \texttt{exatn-mps} backend is selected.

In our implementation of the distributed MPS algorithm each MPI process holds a sub-set of MPS tensors ($N_{qubits}/N_{processes}$). Multiple process groups (ExaTN \texttt{ProcessGroup}) are then created where each process group consists of a pair of neighboring MPI processes to facilitate local communication. The application of entangling gates between neighboring tensors on different MPI processes is performed by:
\begin{itemize}
\item Use \texttt{replicateTensor} API within a pair of neighboring MPI processes to broadcast the MPS tensor right-to-left.
\item The left process (smaller MPI rank) performs the contraction and SVD decomposition locally.
\item The resulting right tensor will then be forwarded left-to-right using \texttt{replicateTensor}. The two processes now decouple and can continue their independent processing of gates on their subset of managed qubits.
\end{itemize}

\subsection{Density Matrix Simulation}

In TNQVM, we can also construct the density matrix by taking the outer product of a state vector with its dual. In this form the density matrix tensor has a rank of 2N, with N being the number of qubits. Using a density matrix representation of the quantum state, we can thus incorporate (non-unitary) noise processes into the simulation workflow. A convenient representation for noise processes (channels) is the Kraus expansion,
\begin{equation}
\rho \mapsto \sum_k A_k \rho A_k^\dagger,
\label{eq:kraus_sum}
\end{equation}
where $\rho = |\Psi\rangle \langle\Psi|$ is the density matrix and $\{A_k\}$ is the set of Kraus operators, satisfying $\sum_k A_k^\dagger A_k=1$, describing the channel of interest.

\label{sec:dm_sim}
\begin{figure}[h!] 
\centering
\begin{subfigure}{.6\textwidth}
  \centering
  \includegraphics[width=0.8\textwidth]{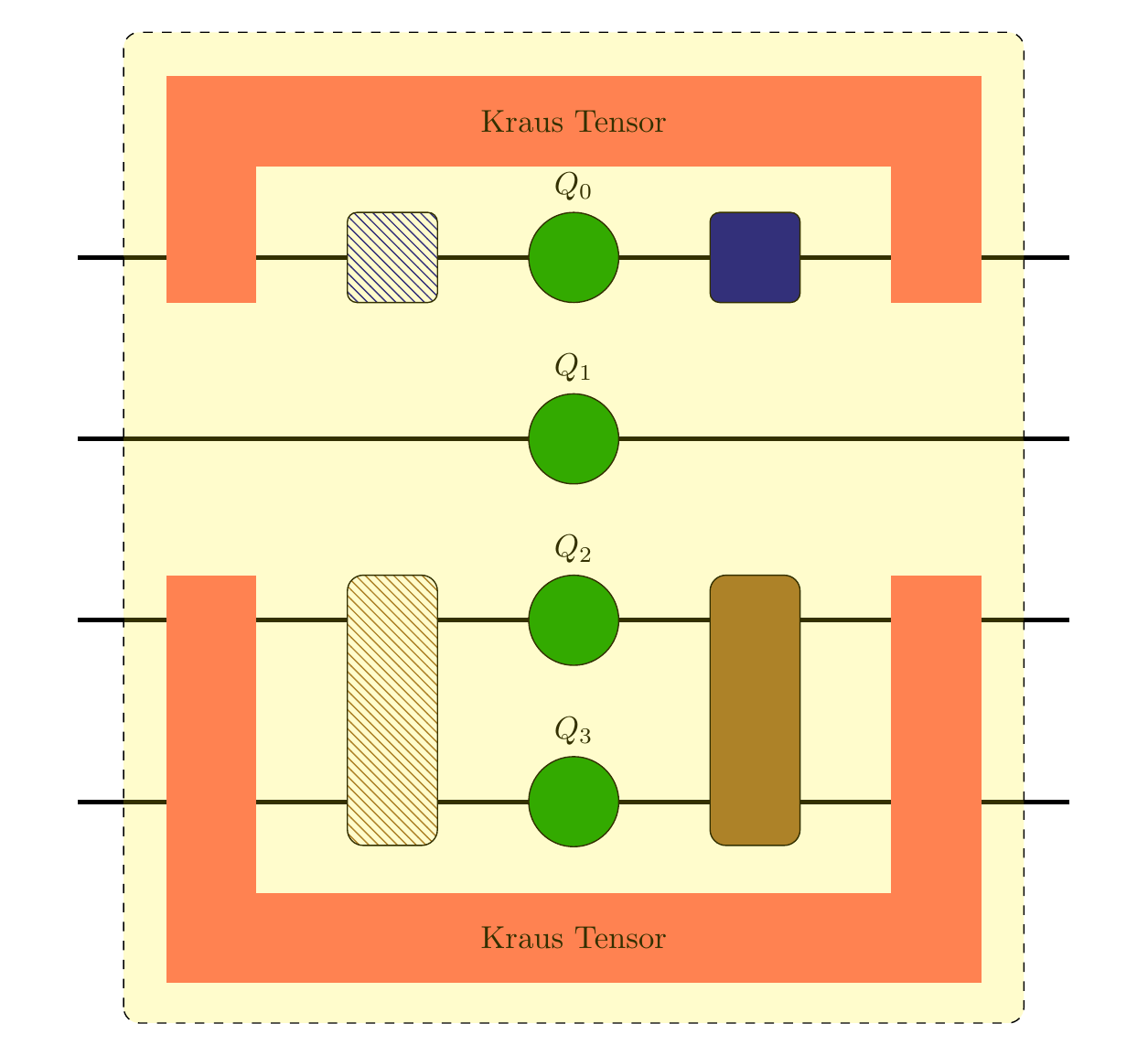} 
  \caption{Cartoon of a density matrix simulation. The circles represent a tensor product of unentangled qubit density matrices, the adjoint action of one and two qubit unitaries is indicated by the rounded rectangles, and the local Kraus superoperators are U-shaped tensors.}
  \label{fig:dm_tensor_net}
\end{subfigure}%
\begin{subfigure}{.4\textwidth}
  \centering
  \includegraphics[width=0.8\textwidth]{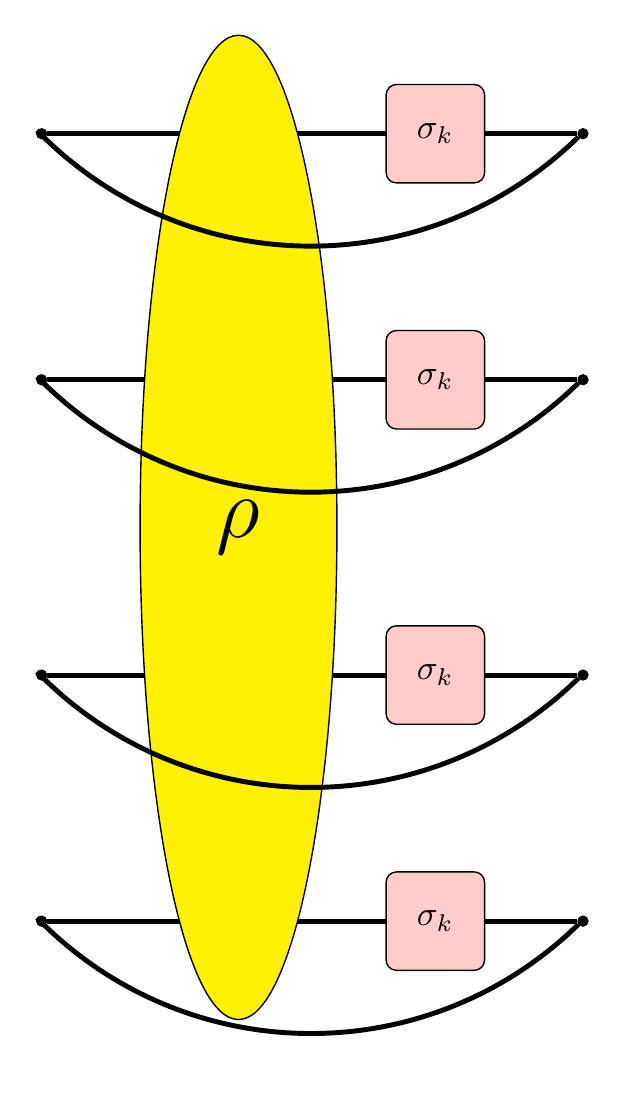} 
  \caption{Expectation value by trace}
  \label{fig:dm_exp_val_trace}
\end{subfigure}
\caption{TNQVM density matrix simulation with noise inclusion}
\label{fig:exatn_dm}
\end{figure}

To simulate noisy quantum circuits via our \texttt{exatn-dm} backend, we append gate tensors to both sides of the tensor network representing the density matrix. Specifically, for each quantum gate, the gate tensor and its conjugate are applied to the ket (right) and bra (left) sides, respectively. Noise operators, on the other hand, are tensors that need to be connected to {\em both} ket and bra legs as shown in Fig.~\ref{fig:dm_tensor_net}. We want to note that for illustration purposes, noise tensors are represented as U-shaped tensors in Fig.~\ref{fig:dm_tensor_net}. We construct them as rank-4 and rank-8 tensors for single- and two-qubit noise processes, respectively, and then append them to our hierarchical tensor network. In our examples, we have defined depolarizing and dephasing Kraus tensors. To formally construct these tensors, we contract (trace over) an environmental qubit in a dilated unitary formulation~\cite{nielsen00}.

Following this construction procedure, we have a full tensor network capturing the noisy evolution according to the input quantum circuit and a given noise model. At this point, we can evaluate this tensor network to retrieve the density matrix, whose diagonal elements equal the probability of measuring a particular computational basis state. For larger systems, however, full density matrix contraction is not practical due to memory constraints. We can compute specific quantities such as bit-string probabilities or expectation values by adding projection or observable tensors and then contracting the bra and ket qubit legs to form a trace value as depicted in Fig.~\ref{fig:dm_exp_val_trace}. The final tensor network is then submitted to ExaTN, which will analyze the structure of the network to determine the tensor contraction sequence and perform the evaluation similar to the simulation algorithm described in Sec.~\ref{sec:direct_contraction}.

Using the \texttt{exatn-dm} backend of TNQVM, users can incorporate quantum noise models into the simulation workflow, such as those that mimic the IBM-Q hardware backends. XACC provides utilities to convert IBM's JSON-based backend configurations into concrete relaxation and depolarization Kraus tensors which are subsequently incorporated into the density matrix tensor network as illustrated in Fig.~\ref{fig:exatn_dm}.

\subsection{Locally-Purified Matrix Product Operator Simulation}
\label{sec:lmps}
\begin{figure}[h!]
\centering
\begin{subfigure}{.4\textwidth}
  \centering
  \includegraphics[width=0.9\textwidth]{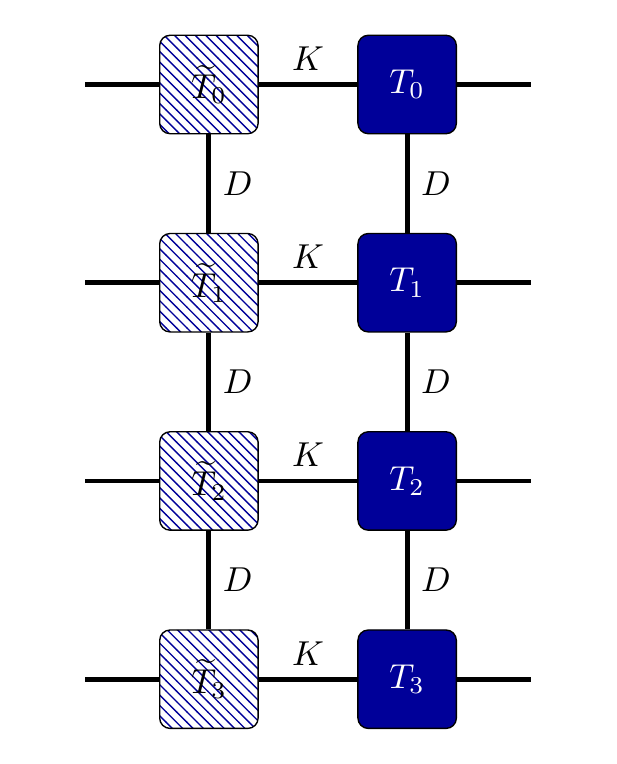} 
  \caption{Locally-purified MPS tensor network: $K$ is the Kraus dimension and $D$ is the bond dimension.}
  \label{fig:pmps_network}
\end{subfigure}
\hspace{12pt}
\begin{subfigure}{.4\textwidth}
  \centering
  \includegraphics[width=0.9\textwidth]{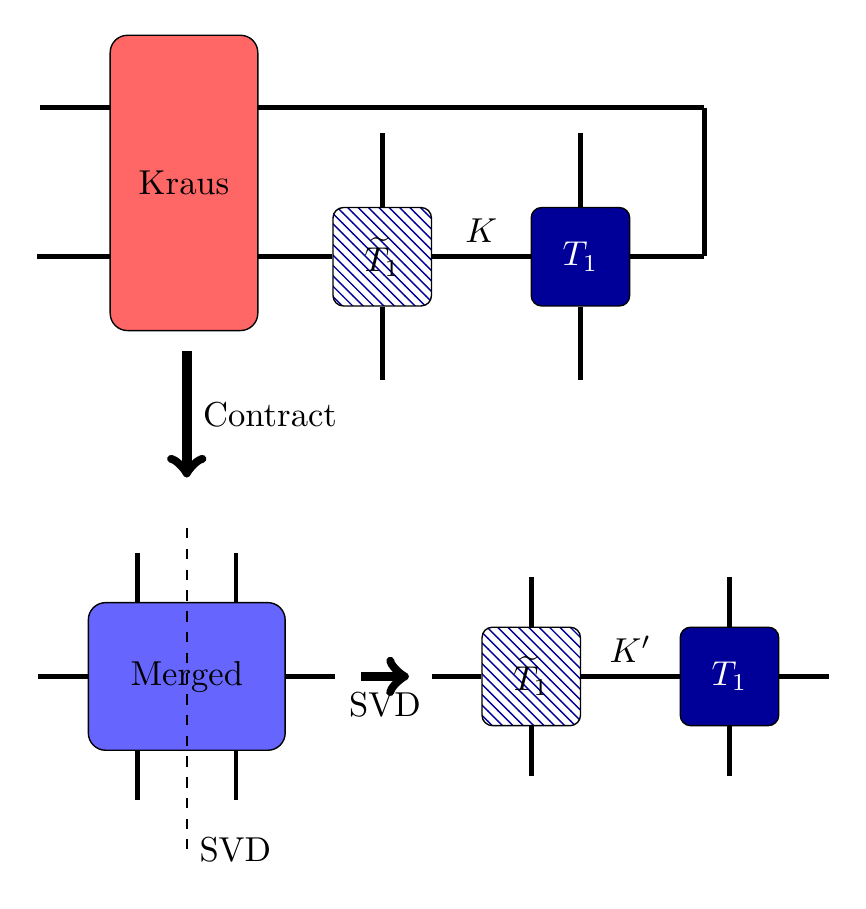} 
  \caption{Contract a Kraus tensor and decompose (SVD) back to the locally-purified product state form. The Kraus dimension is updated ($K$ to $K'$ after the SVD decomposition.}
  \label{fig:pmps_kraus_contract}
\end{subfigure}
\caption{Locally-purified matrix product state tensor network representation.}
\label{fig:exatn_pmps}
\end{figure}
\noindent
Just as one can factorize a full state vector into an MPS tensor network, one can also factorize the full density matrix tensor (as described in Sec.~\ref{sec:dm_sim}) into a similar tensor train structure. One method, known as the locally-purified matrix product state (PMPS) ansatz~\cite{werner2016positive}, is depicted in Fig.~\ref{fig:pmps_network}. It is implemented in TNQVM via another tensor processing visitor backend, named \texttt{exatn-pmps}, that performs approximate density matrix-based simulation following this decomposition procedure.

\begin{wrapfigure}{r}{.48\textwidth}
%\vspace{-40pt}
  \lstset {language=C++}
  \begin{lstlisting}
#include "xacc.hpp"

int main(int argc, char **argv) {
  // Initialize the XACC Framework
  xacc::Initialize(argc, argv);

  // Use ExaTN based TNQVM Accelerator 
  auto qpu = 
    xacc::getAccelerator("tnqvm:exatn", 
                    {{"shots", 1024}});

  // Create a Program
  auto xasmCompiler = 
        xacc::getCompiler("xasm");
  auto ir = xasmCompiler->compile(
  R"(__qpu__ void Bell(qbit q) {
    H(q[0]);
    CX(q[0], q[1]);
    Measure(q[0]);
    Measure(q[1]);
  })", qpu);
  auto program = ir->getComposite("Bell");
  // Allocate a register of 2 qubits
  auto qubitReg = xacc::qalloc(2);
  // Execute
  qpu->execute(qubitReg, program);
  // Print the result in the buffer.
  qubitReg->print();

  // Finalize the XACC Framework
  xacc::Finalize();

  return 0;
}
\end{lstlisting}
\caption{Code snippet demonstrating TNQVM usage with XACC.}
%Reduce white space
\vspace{-50pt}
\label{fig:tnqvm_example_xacc}
\end{wrapfigure}

The application of quantum gates is similar to the MPS backend, as described in Sec.~\ref{sec:mps}. Two-qubit gates are contracted with PMPS tensors, denoted by $T_i$, to form a merged tensor which is then decomposed into the canonical tensor product. This procedure only modifies the virtual bond dimension $D$ (see Fig.~\ref{fig:pmps_network} between the neighboring PMPS tensors) which captures the systems entanglement properties. To simulate a non-unitary channel Kraus tensors, such as the ones shown in Fig.~\ref{fig:dm_tensor_net}, are contracted with the qubit legs of the MPS tensor and its conjugate, see Fig.~\ref{fig:pmps_kraus_contract}. After this contraction, we can apply SVD along the Kraus dimension to recover the canonical (locally-purified) form of the PMPS factorization. The Kraus dimension ($K$) between the MPS tensor and its conjugate, encoding statistical mixtures of the density operator, is updated after each noise operator iteration.

\section{Demonstrations}
\label{sec:demo}
In this section, we seek to demonstrate the utility, flexibility and performance of the various ExaTN-based backends implemented in TNQVM. These demonstrations cover applications from ubiquitous quantum circuit simulation to highly-customized amplitude or bit-string sampling experiments.

\subsection{Quantum circuit simulation}
As a first example, Fig.~\ref{fig:tnqvm_example_xacc} shows a typical usage of TNQVM as a virtual Accelerator in the XACC framework. In particular, after TNQVM is compiled and installed to the XACC plugin directory, users can use \texttt{xacc::getAccelerator} API to retrieve an instance of the TNQVM accelerator using the name key \texttt{tnqvm}. In addition, one of the backends described in Section 3 can be specified after the ':' symbol. For example, the code snippet in Fig.~\ref{fig:tnqvm_example_xacc} calls for the full tensor network contraction simulator (\texttt{exatn}).

Any specialized configurations are given in terms of a dictionary (key-value pairs) when requesting the accelerator. For example, we can specify the number of simulation runs (shots), as shown in Fig.~\ref{fig:tnqvm_example_xacc}. There are a lot of configurations specific to each simulator backend documented on the XACC documentation website. Simulation results, e.g., shot count distribution, are persisted to the qubit register (\texttt{xacc::AcceleratorBuffer}) for later retrieval or post-processing.

\begin{wrapfigure}{l}{.55\textwidth}
\lstset {language=C++}
\begin{lstlisting}
// Query the noise model from an IBM device
auto noiseModel = 
    xacc::getService<xacc::NoiseModel>("IBM");
noiseModel->initialize(
            {{"backend", "ibmqx2"}});
auto qpu = xacc::getAccelerator(
    "tnqvm:exatn-dm", 
    {{"noise-model", noiseModel}});

auto qubitReg = xacc::qalloc(1);

// Create a test program: 
// Apply back-to-back Hadamard gates 
// to assess gate noise
auto xasmCompiler = xacc::getCompiler("xasm");
auto ir = xasmCompiler->compile(R"(
__qpu__ void conjugateTest(qbit q) {
    for (int i = 0; i < NB_CYCLES; i++) {
      H(q[0]);
      H(q[0]);
    }
    Measure(q[0]);
  })", qpu);
\end{lstlisting}
\caption{Noisy quantum circuit simulation with TNQVM. The device noise model (IBMQ Yorktown, \texttt{ibmqx2}) is generated from online calibration data and provided to TNQVM as a \texttt{noise-model} configuration. In this code snippet, we show the experiment on the first qubit (\texttt{q[0]}) of the device. Other qubits can also be experimented with similarly by specifying their indices. }
\label{fig:tnqvm_noisy}
\vspace{-20pt}
\end{wrapfigure}

The above example demonstrates the seamless integration of TNQVM and all of its backends into the XACC stack. All user codes can use TNQVM as a drop-in replacement for the backend Accelerator. Furthermore, when the simulation demands an HPC platform, users will get instant scalability, i.e., no code changes required, thanks to the TNQVM-ExaTN abstraction layer.

\subsection{Noisy simulation}
One of the advantages of being part of the XACC framework is that TNQVM can query device characteristics of hardware backends, e.g., IBMQ devices, from XACC to perform hardware emulation. Since TNQVM fully supports noisy quantum circuit simulations, local noise channels can be incorporated into the simulation process. In Fig.~\ref{fig:tnqvm_noisy}, we show a simple example how one can construct a noise model from the IBMQ \texttt{ibmqx2} (Yorktown) device configuration via the XACC \texttt{NoiseModel} utility, followed by the initialization of the density matrix based backend of TNQVM (\texttt{exatn-dm}, see Sec.~\ref{sec:dm_sim}).

\begin{wrapfigure}{r}{.6\textwidth}
\vspace{-20pt}
\begin{tikzpicture}
\begin{axis}[
      cycle list name=exotic,
      legend columns=3,
      xmin = 0, xmax = 700,
      xlabel = {Sequence length (\texttt{NB\_CYCLES})}, 
      ylabel = {$\langle Z_i \rangle$}, 
      y label style={at={(axis description cs:0.1,0.5)},anchor=south},
      title = {Benchmarking of noisy Hadamard gates with TNQVM}]
\addplot
table[x=x, y=y] {
x  y
1  0.999457
%2  0.998915
%3  0.998373
%4  0.997831
%5  0.99729
%6  0.996749
%7  0.996208
%8  0.995668
%9  0.995127
10  0.994587
%11  0.994048
%12  0.993508
%13  0.992969
%14  0.992431
%15  0.991892
%16  0.991354
%17  0.990816
%18  0.990278
%19  0.989741
20  0.989204
%21  0.988667
%22  0.988131
%23  0.987595
%24  0.987059
%25  0.986523
%26  0.985988
%27  0.985453
%28  0.984918
%29  0.984384
30  0.98385
%31  0.983316
%32  0.982782
%33  0.982249
%34  0.981716
%35  0.981184
%36  0.980651
%37  0.980119
%38  0.979587
%39  0.979056
40  0.978525
%41  0.977994
%42  0.977463
%43  0.976933
%44  0.976403
%45  0.975873
%46  0.975343
%47  0.974814
%48  0.974285
%49  0.973757
50  0.973228
%51  0.9727
%52  0.972172
%53  0.971645
%54  0.971118
%55  0.970591
%56  0.970064
%57  0.969538
%58  0.969012
%59  0.968486
60  0.96796
%61  0.967435
%62  0.96691
%63  0.966386
%64  0.965861
%65  0.965337
%66  0.964814
%67  0.96429
%68  0.963767
%69  0.963244
70  0.962721
%71  0.962199
%72  0.961677
%73  0.961155
%74  0.960633
%75  0.960112
%76  0.959591
%77  0.959071
%78  0.95855
%79  0.95803
80  0.95751
%81  0.956991
%82  0.956472
%83  0.955953
84  0.955434
100 0.947173
200 0.897137
300 0.849744
400 0.804855
500 0.762337
600 0.722065
700 0.68392
};
\addlegendentry{Q0};

\addplot
table[x=x, y=y] {
x  y
0   1.0
10  0.971364
20  0.943547
30  0.916528
40  0.890282
50  0.864787
60  0.840023
70  0.815968
80  0.792601
90  0.769904
100 0.747857
200 0.55929
300 0.418269
400 0.312805
500 0.233934
600 0.174949
700 0.130837
};
\addlegendentry{Q1};
\end{axis}
\end{tikzpicture} 
\caption{Plots of expectation values of Pauli-Z operator vs the length of the H-H sequence (\texttt{NB\_CYCLES} in Fig.~\ref{fig:tnqvm_noisy}). Since TNQVM simulates all noise channels according to the device model (\texttt{ibmqx2}), the $\langle Z \rangle$ expectation decays as the number of cycles increases. Calibration data: Single qubit Pauli-X error: 8.906e-4 (Q0) and 1.935e-3 (Q1).}
\label{fig:h_gate_chart}
\vspace{0pt}
\end{wrapfigure}
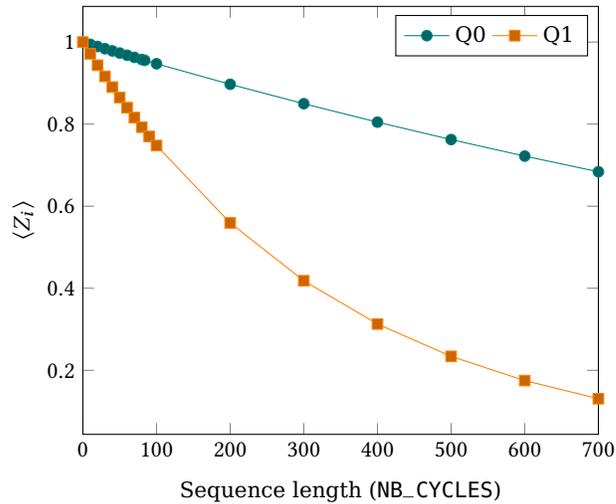

In this demonstration, we simulate a simplified randomized benchmarking procedure whereby the gate set only contains a single gate (Hadamard). By repeating this gate back-to-back over multiple cycles, we can quantify the gate noise in terms of deviation from an ideal identity operation. In other words, if the Hadamard gate is ideal, we would see the qubit state stays at 0 ($\langle Z \rangle = 1$) regardless of the number of gates. However, due to device noise, we expect a decay of the ground state population as the number of cycles increases, and we see this in the resultant data shown in Fig.~\ref{fig:h_gate_chart}.

In this experiment, we test the Hadamard gate sequence on both qubit 0 and 1, which have a quite significant single-qubit gate fidelity difference (8.906e-4 for Q0 and 1.935e-3 for Q1). It is worth noting that these calibration parameters are provided by IBM in real-time, which XACC uses to construct the \texttt{NoiseModel} object. The simulation results from TNQVM, as shown in Fig.~\ref{fig:h_gate_chart}, are consistent with the device characteristics. We can clearly see a much faster decay for Q1, whose gate error rate is reported to be more than double that of Q0.

Using the matrix trace bra-ket connection, as depicted in Fig.~\ref{fig:dm_exp_val_trace}, we can simulate noisy quantum circuits that contain a large number of qubits as low-rank tensor networks. Intermediate tensor slices appearing in these large-scale tensor network contractions can be effectively distributed across many compute nodes by ExaTN.

\subsection{Single amplitude calculation}
In order to demonstrate the parallel performance and efficient use of GPU by the TNQVM-ExaTN simulator, here we examine the run time of the direct tensor contraction algorithm when simulating a single output state amplitude (see Table~\ref{table:exatn_utils}) for the Sycamore random quantum circuits involving 53 qubits~\cite{arute2019quantum}. The code snippet setting up the simulation experiment is shown in Fig.~\ref{fig:tnqvm_sycamore_amplitude}, whereby we can recognize the familiar \texttt{Accelerator} initialization,  \texttt{AcceleratorBuffer} allocation, and execution workflow patterns.
The only difference is that we provide a bit-string initialization parameter to request that the amplitude of that specific bit-string be computed.

The Sycamore test circuits involve a large number of qubits (53), thus making the full state-vector calculation unfeasible. Instead, we use the projection, as shown in Fig.~\ref{fig:ampl_calc}, to compute the amplitude of a particular bit-string of interest. 
\begin{figure}[ht!]
  \lstset {language=C++}
  \begin{lstlisting}
// Compute the amplitude of a bit-string:
// BIT_STRING is a vector of length 53, e.g. 000000000...00
// represents the state whose amplitude we want to compute.
auto qpu =
    xacc::getAccelerator("tnqvm:exatn", {{"bitstring", BIT_STRING},
                                         {"exatn-buffer-size-gb", 2}});
// Allocate a register of 53 qubits (Sycamore chip)
auto qubitReg = xacc::qalloc(53);
// Program is the random quantum circuit.
qpu->execute(qubitReg, program);
// Retrieve the amplitude result
const double realAmpl = (*qubitReg)["amplitude-real"].as<double>();
const double imagAmpl = (*qubitReg)["amplitude-imag"].as<double>();
\end{lstlisting}
\caption{Simulating a Sycamore bit-string amplitude with TNQVM. Variable \texttt{program} is an XACC's \texttt{CompositeInstruction} instance compiled from the Sycamore test circuits.}
\label{fig:tnqvm_sycamore_amplitude}
\end{figure}
Also, since we intended to run this test on a cluster, configurations such as the RAM buffer size per MPI process can be customized when initializing the TNQVM accelerator. The random quantum circuit (\texttt{program} variable in Fig.~\ref{fig:ampl_calc}) is adopted from the Google's quantum supremacy experiment~\cite{arute2019quantum} in which the ideal simulation of the depth-14 circuit on Summit was already considered prohibitively expensive at that time. The performance results for the depth-14 Sycamore random quantum circuit are listed in Table~\ref{table:benchmark-data}. Our compilation of the circuit comprises 2828 quantum gates, a higher count than originally reported because some 1-body gates had to be additionally decomposed inside XACC, which does not affect computational complexity (the number of two-body gates is the same).

\begin{table}[ht!]
\caption{Performance comparison of simulating a single amplitude of the depth-14 2D random quantum circuit with 53 qubits.}
\label{table:benchmark-data}
 \begin{tabular}{ p{0.25\textwidth}  p{0.07\textwidth} p{0.15\textwidth} p{0.12\textwidth} p{0.12\textwidth} p{0.12\textwidth}} 
 System & Precision  & \multicolumn{1}{p{0.15\textwidth}}{\centering Time to solution [s]} & \multicolumn{1}{p{0.12\textwidth}}{\centering Avg. Tflop/s/GPU} & \multicolumn{1}{p{0.12\textwidth}}{\centering Flop count per GPU} & \multicolumn{1}{p{0.12\textwidth}}{\centering Bit-string amplitude} \\ 
  \hline
  \hline
  \multirow{2}{0.25\textwidth}{DGX-A100, 8 A100 GPU} & \multicolumn{1}{l}{FP32} &	\multicolumn{1}{l}{2003.23}	& \multicolumn{1}{l}{15.06}	& \multicolumn{1}{l}{3.0160E+16} & \multicolumn{1}{l}{6.4899E-09} \\
  & \multicolumn{1}{l}{TF32} & \multicolumn{1}{l}{868.38} & \multicolumn{1}{l}{34.73} & \multicolumn{1}{l}{3.0160E+16} & \multicolumn{1}{l}{6.4840E-09} \\ 
  \hline

DGX-1, 8 V100 GPU &	FP32 &	13028.92 &	3.05 &	3.9791E+16 &	6.4896E-09 \\
\hline
\textbf{OLCF Summit} & {} &	{}	& {} &	{}	& {} \\
16 nodes, 96 V100 GPU &	FP32 &	695.5 &	4.03 &	2.7995E+15	& 6.4899E-09 \\
64 nodes, 384 V100 GPU & FP32 &	125.52 &	4.85 &	6.0856E+14 &	6.4899E-09 \\ 
64 nodes, 384 V100 GPU\tablefootnote{Faster tensor network contraction path} &	FP32 &	60.695	& 7.99 &	4.8508E+14	& 6.4900E-09 \\
\hline
\textbf{Dual AMD Rome CPU} & {} &	{}	& {} &	{}	& {} \\
1 node, 2 x 64-core CPU &	FP32 &	{40571.41\tablefootnote{Extrapolated after 2-hour execution}} & {2.98} &	{3.0160E+16} & {6.4899E-09} \\
\hline
\hline

 %4 & 877 & 1 & ?? & {}\\
 %{} & {} & 2 & ?? & 2x??\\
 %{} & {} & 4 & ?? & 4x??\\
% \hline
 %12 & 12 & 72 & 22102.17 & 317.4 & 5.802 \\
 %{}& 64 & 384 & 15427.92 & 25.0 & 9.63\\
 %\hline
 %14 & 64 & 384 & 805929.31 & 1035.0 & 12.17  \\
\end{tabular}
\end{table}

As seen from Table~\ref{table:benchmark-data}, on Summit~\footnote{Summit node: 2 IBM Power9 CPU with 21 cores each, 6 NVIDIA V100 GPU with 16 GB RAM each, NVLink-2 all-to-all, 512 GB Host RAM} we observe an excellent strong scaling (from 16 to 64 nodes) as well as a reasonably good absolute efficiency (27 - 53\% of the theoretical FP32 peak per GPU). The second 64-node experiment on Summit used a faster tensor contraction path which took longer to find as the price to pay. The 16-node and the first 64-node experiments spent less time in the tensor contraction path search than in its actual execution, whereas the second 64-node experiment spent more time in finding a faster tensor contraction path than in its actual execution (this contraction path also turned out to deliver a better Flop efficiency). In all these experiments we used an out-of-core tensor contraction algorithm implemented in ExaTN, in which the participating tensors may exceed the GPU RAM limit as long as they fit in a normally larger Host RAM. The performance of this algorithm can easily become bound by the Host-to-Device data transfer bandwidth that can be clearly seen from the DGX-1~\footnote{DGX-1 node: 2 Intel Xeon E5-2698 CPU with 20 cores each, 8 NVIDIA V100 GPU with 32 GB RAM each, PCIe-3 bus between CPU and GPU, 512 GB Host RAM} results where GPUs communicate with the CPU Host via the slower PCIe-3 bus instead of faster NVLink-2. Combined with a lesser amount of Host RAM per MPI process (8 MPI processes on 8 V100 GPUs versus 6 MPI processes on 6 V100 GPUs on Summit), it resulted in a significant performance drop, down to about 20\% of the absolute FP32 peak per GPU. On the other hand, the new DGX-A100 box~\footnote{DGX-A100 node: 2 AMD Rome CPU with 64 cores each, 8 NVIDIA A100 GPU with 80 GB RAM each, PCIe-4 bus between CPU and GPU, 2 TB Host RAM} with 2 TB of Host RAM and 80 GB RAM per GPU, as well as with a faster PCIe-4 bus, delivers much better FP32 performance for the out-of-core algorithm. Furthermore, the new A100 tensor cores running with the TF32 precision bump up the performance with an impressive additional 2.3X speed up while keeping the result correct to 3 decimal digits. Despite such a great performance of the out-of-core algorithm on DGX-A100 with a quickly generated, but suboptimal tensor contraction sequence, the simulation of the Sycamore random quantum circuits with a higher depth will result in a computational workload with a lower arithmetic intensity, necessitating the full transition of all tensors into the GPU RAM (in-core), which we plan to enable soon.

\subsection{Marginal wave-function slice calculation}
Similar to the bit-string amplitude calculation (Fig.~\ref{fig:tnqvm_sycamore_amplitude}), one can also use TNQVM to compute a marginal wave-function (state-vector) slice for a subset of qubits given other qubits are projected to classical 0 or 1 states. In particular, by keeping a set of $n$ qubits open, the computed vector slice will have a length of $2^n$, representing the marginal (conditional) wave function of these qubits given other qubits are fixed. This calculation applies to the chaotic sampling of random quantum circuits or divide-and-conquer parallelization of the state-vector based simulation.

\begin{wrapfigure}{r}{.49\textwidth}
\vspace{-20pt}
  \lstset {language=C++}
  \begin{lstlisting}
// Create a Program
auto program = xasmCompiler->compile(
R"(__qpu__ void entangle(qbit q) {
    H(q[0]);
    for (int i = 1; i < 60; i++) {
        CX(q[0], q[i]);
    }
})")->getComposite("entangle");

const int NB_QUBITS = 60;

// Measure two random qubits
// Compute the marginal wavefunction slice
// of two random qubits (Q2 and Q47)
BIT_STRING[2] = -1;
BIT_STRING[47] = -1;
// Note: Other bits in the BIT_STRING 
// array can be set to 0 or 1
// to denote their projection values.
auto qpu = 
    xacc::getAccelerator("tnqvm:exatn", 
            {{"bitstring", BIT_STRING}});

// Allocate qubit register and execute
auto qubitReg = xacc::qalloc(NB_QUBITS);
qpu->execute(qubitReg, program);
\end{lstlisting}
\caption{Compute the marginal wave function slice for a subset of qubits (2) out of a qubit register of size 60. The qubits are initialized to a cat state.}
\label{fig:tnqvm_wave_fn_slice}
\vspace{-30pt}
\end{wrapfigure}

Fig.~\ref{fig:tnqvm_wave_fn_slice} demonstrates the use of this calculation for a many-qubit cat state, namely, the state $|\Psi\rangle = \frac{1}{\sqrt{2}}(|00..00\rangle + |11..11\rangle)$.
% \begin{equation}
% |\Psi\rangle = \frac{1}{\sqrt{2}}(|00..00\rangle + |11..11\rangle)
% \label{eq:cat_state}
% \end{equation}
This state is generated by the Hadamard gate followed by a sequence of entangling CNOT gates. For demonstration purposes, we randomly selected two open qubits, 2 and 47, to compute the wave-function slice. Thus, the result vector is expected to have a length of 4 and will be returned in the \texttt{AcceleratorBuffer}.

In TNQVM, we use a special value of -1 to denote open qubits in the input bit-string. Other qubits are projected to either 0 or 1 values as specified in the bit-string. For instance, given the cat-state, only when other qubits are all 0's or all 1's, then the marginal wave-function result is non-zero. And, we get a marginal state vector of [1,0,0,0] or [0,0,0,1] for the two cases of all others are 0's or all 1's, respectively. This confirms the expected entanglement property of the cat state. 
As reported in Ref.~\cite{pan2021simulating}, we can further draw bit-string samples from the resulting marginal wave-function slice to simulate sampling from a subspace of the total Hilbert space determined by those projected qubits.

\section{Conclusion}
\label{sec:conclusion}
\par
We have introduced a general tensor network based quantum circuit simulator capable of modeling both ideal and noisy quantum circuits as well as computing various experimentally accessible properties depending on the tensor network formalism used. The versatility and scalability of the ExaTN numerical backend enables TNQVM to simulate large-scale quantum circuits efficiently on leadership HPC platforms. In addition, we have also demonstrated an algorithm that incorporates noisy dissipative processes into the simulation.
\par
TNQVM provides a number of capabilities allowing users to efficiently calculate expectation values, exactly or approximately, generate unbiased random measurement bit-strings, or compute state-vector amplitudes. These properties are pertinent to near-term experimental endeavors, such as studying the variational quantum algorithms and validating the new quantum hardware. In this respect, TNQVM presents itself as a valuable tool for analysis and verification of quantum algorithms and devices in pursuit of advancing the progress towards large-scale, fault-tolerant quantum computing. Our continuous goal is to keep extending the functionality of TNQVM and ExaTN by incorporating new and more efficient tensor-based techniques into the simulation workflow in order to enable classical simulation of larger and deeper quantum circuits.

\section*{Acknowledgments}
This work has been supported by the US Department of Energy (DOE) Office of Science Advanced Scientific Computing Research (ASCR) Quantum Computing Application Teams (QCAT), Accelerated Research in Quantum Computing (ARQC), and National Quantum Information Science Research Centers. The development of the core capabilities of the ExaTN library was funded by a laboratory directed research and development (LDRD) project at the Oak Ridge National Laboratory (LDRD award 9463). This research used resources of the Oak Ridge Leadership Computing Facility, which is a DOE Office of Science User Facility supported under Contract DE-AC05-00OR22725. Oak Ridge National Laboratory is managed by UT-Battelle, LLC, for the US Department of Energy under contract no. DE-AC05-00OR22725. We would also like to thank Tom Gibbs and NVIDIA for providing access to the DGX-A100 computational resources for performance benchmarking.

\bibliographystyle{plain}
\bibliography{main}
\end{document}